\documentclass[iop]{emulateapj} 

\usepackage{graphicx}

\usepackage{amssymb}
\usepackage{tabto}

\usepackage{epsfig}
\usepackage{amsmath}
\usepackage{xspace}

\usepackage{ifpdf}
\usepackage{calc}
\usepackage{comment}

\newcommand{\swift}{{\it Swift}\xspace}
\newcommand{\nustar}{{\it NuSTAR}\xspace}

\newcommand{\pe}[1]{{\it ProtoEXIST{#1}}\xspace}

\newcommand{\sS}[1]{\mbox{$\rm{}^{#1}$}}

\newcommand{\Deg}{\mbox{$^\circ$}\xspace}
\newcommand{\x}{\mbox{$\times$}\xspace}

\begin{document}




\title{Imaging Analysis of the Hard X-ray Telescope \pe2\ \ and
New Techniques for High Resolution Coded Aperture Telescopes
}


\author{
Jaesub Hong\altaffilmark{1}, 
Branden Allen\altaffilmark{1}, 
Jonathan Grindlay\altaffilmark{1},
Scott Barthelmy\altaffilmark{2}
}
\altaffiltext{1}{Harvard-Smithsonian Center for Astrophysics, Cambridge, MA 02138}
\altaffiltext{2}{NASA Goddard Space Flight Center, Greenbelt, MD 20771}

\begin{abstract}

Wide-field ($\gtrsim$ 100 deg\sS{2}) hard X-ray coded-aperture telescopes
with high angular resolution ($\lesssim$ 2$'$) will enable a wide range
of time domain astrophysics.  For instance, transient sources such
as gamma-ray bursts can be precisely localized without assistance of
secondary focusing X-ray telescopes to enable rapid followup studies.
On the other hand, high angular resolution in coded-aperture imaging
introduces a new challenge in handling the systematic uncertainty: average
photon count per pixel is often too small to establish a proper background
pattern or model the systematic uncertainty in a time scale where the
model remains invariant.  We introduce two new techniques to improve
detection sensitivity, which are designed for, but not limited to high
resolution coded-aperture system: a self-background modeling scheme which
utilizes continuous scan or dithering operations, and a Poisson-statistics
based probabilistic approach to evaluate the significance of source
detection without subtraction in handling the background.  We illustrate
these new imaging analysis techniques in high resolution coded-aperture
telescope using the data acquired by the wide-field hard X-ray telescope
\pe2 during the high-altitude balloon flight in Fall, 2012.  We review
the imaging sensitivity of \pe2 during the flight, and demonstrate
the performance of the new techniques using our balloon flight data in
comparison with simulated ideal Poisson background.
\end{abstract}

\keywords{hard X-ray Imaging --- CdZnTe --- Coded-aperture imaging}

\section{Introduction}

Near-arcmin angular resolution for wide-field\footnote{Here we loosely define
wide field as a solid angle $\gtrsim$ 100 deg\sS{2}, which enables
all sky survey in a reasonable time scale and depth. For instance, a
telescope with a 100 deg\sS{2} field of view can scan
the entire sky in roughly 2 yr with an average exposure of 100 ks.} 
hard X-ray coded-aperture telescopes is within reach in near future.
The Burst Alert Telescope (BAT) on \swift, which have been
operating successfully for more than a decade, covers a 1.4 sr
field (50\% coding\footnote{Field of views of coded-aperture
systems quoted in this paper refer to 50\% coding unless otherwise
noted.}) with the 22\arcmin\ angular resolution in the
15--150 keV band \citep{Gehrels04}.
The Imager on Board the {\it INTEGRAL} Satellite 
(IBIS) and the Joint European X-ray Monitor (JEM-X) can observe a $\sim$
19 \x19 deg\sS{2} field with the 12\arcmin\ resolution in the 15 keV --
10 MeV band and a 7.5\Deg diameter field with the 3\arcmin\ 
resolution in the 3--35 keV band, respectively \citep{Winkler03}.

Recently we have achieved the 5\arcmin\ angular
resolution over a 20 \x 20 deg\sS{2} field in the 5 --
200 keV band with a
balloon-borne hard X-ray coded-aperture telescope \pe2 \citep{Hong13}
by employing an array of CdZnTe detectors with the
Application Specific Integrated Circuits (ASICs) of a high pixel
density used in Nuclear Spectroscopic Telescope Array (\nustar)
\citep{Harrison13}.  Next generation ASICs with a higher pixel density
will soon enable $\lesssim$ 2\arcmin\ resolution, which will allow
source localization within 20\arcsec. High precision source
localization from a wide-field hard X-ray telescope can initiate
rapid follow-up studies of transient sources with many 
telescopes around the world without assistance of secondary focusing
X-ray telescopes, and thus open a wide range of discovery
space in the time domain astrophysics.

High resolution in coded-aperture telescopes introduces a new challenge
in handling the systematics such as non-uniform detector background. Even
with a decent exposure, observed photon counts per pixel often remain too
low to establish the precise pattern of non-uniformity in the detector
plane, which can limit the detection and localization sensitivity. Novel
observing schemes such as continuous scan or dithering
can reduce the unknown systematic errors \citep{Grindlay04}, but as the pixel
density increases, additional care in the analysis is required to
ensure the high sensitivity.
Here we introduce two new techniques to alleviate the effects of
systematics and improve detection sensitivity
even under low count statistics of high resolution
coded-aperture telescopes.
We illustrate these
new imaging analysis techniques 
using the data acquired by the wide-field high resolution 
hard X-ray telescope \pe2 during the high-altitude balloon flight in Fall, 2012. 

In Section \ref{s:instr}, we review the basic parameters of the \pe2
telescope.  In Section \ref{s:aspect},
we describe the boresight calibration and the performance of our pointing
and aspect system during the high altitude balloon flight in Fall, 2012.  In
Section \ref{s:sens}, we estimate the sensitivity limit of the \pe2
telescope and illustrate the challenge of high resolution coded-aperture
telescopes using the \pe2 observations during
the flight.  In Section \ref{s:sbc}, we introduce a self-correcting
background modeling scheme, which utilizes continuous scan or
dithering operations.  In Section \ref{s:tm}, we introduce a Poisson statistics
based detection significance map called `trial map' \citep{Hong16} to
coded-aperture imaging, which can handle the background without subtraction.

\begin{table}
\small
\caption{Telescope Parameters of \pe2}
\label{t:telpar}
\begin{tabular*}{0.470\textwidth}{r@{\extracolsep{\fill}}l}
\hline\hline
Parameters			&	Values	\\
\hline
Sensitivity			&	$\sim$ 140 mCrab/hr\sS{a}	\\
Energy Range			&	5 -- 200 keV			\\
Energy Resolution		&	2 -- 3 keV			\\
Field of View 			&	20\Deg \x 20\Deg (50\% Coding)	\\
Angular Resolution		&	4.8$'$	\\
\hline
CZT Detector			&	56 crystals \x (1.98 \x 1.98 cm\sS{2}) \\
Active Area 			&	220 cm\sS{2}			  \\
Pixel, Thickness		&	0.6 mm, 5 mm	\\
\hline
Tungsten Mask 			&	4 layers \x 0.1 mm thick	\\
Coding Area			&	33.3 \x 33.3 cm\sS{2} 	\\
Pixel, Grid, Thickness		&	1.1 mm, 0.1 mm, 0.4 mm	\\
Pattern				&	Random	\\
\hline
Mask-Det. Separation		&	90 cm		\\				
Rear and Side Shields		&	{Graded Pb/Sn/Cu} \\
\sS{241}Am Cal. Source		&	{220 nCi each, $\sim$ 36 cm above det}	\\
\hline
\end{tabular*}\\
(a) Without atmospheric absorption. The atmospheric absorption at altitude of 40 km
can reduce the signal by a factor of $\sim$ 3 in the 30--100 keV band, which depends
on the source spectrum and the pointing elevation.
\end{table}

\section{\pe2} \label{s:instr}


The \pe\ payload during the flight in 2012 consisted of the two X-ray
telescopes - \pe1 \& {\it 2} and a day-time optical star camera for pointing
guidance and aspect correction.  The detailed description of the
instruments and the flight performance of
the detector system are found in \citet{Hong13}.  
In this paper, we focus on the imaging performance of the \pe2\  telescope.
Table~\ref{t:telpar} summarizes the telescope parameters of 
\pe2. 

The \pe2\ telescope is a wide-field hard X-ray telescope with
an array of CZT detectors and a Tungsten mask. High pixel density in
the CZT detectors and the Tungsten mask enables 5$'$ angular resolution over
more than 20\Deg \x 20\Deg field-of view (FoV) of 50\% coding fraction.
The combined thickness of 0.4 mm in the Tungsten mask can modulate
X-rays up to $\sim$ 200 keV. Low noise ASICs used in \nustar and 
thick CZT crystals (5 mm) cover a wide energy range from 5 to 200 keV, 
although at a typical flight altitude ($\sim$ 40 km) X-rays below $\sim$ 30 keV
from celestial sources are undetectable due to absorption in the
remaining atmosphere along the line of sights.

\subsection{Pointing \& Aspect System}\label{s:aspect}


 A few days before the flight in 2012, we measured the boresight
offset between the X-ray telescopes and the optical star camera using 
an \sS{241}Am radioactive source \citep{Hong13}. The analysis showed that
a relative offset is ($\Delta$az, $\Delta$el) = (+5$'$,+39$'$) for \pe2
when there is no pressure difference between the inside and the outside
of the pressure vessel (PV).  
During the flight, the pressure difference between the inside and the outside
of the PV forced the top cap of the PV to bow
out by a few degrees.  
Since the coded masks were mounted on the top
cap, this introduced an additional shift between the X-ray telescopes
and the star camera. We measured the additional offset by creating an
artificial pressure difference on the ground
after the flight. Under an about 12 psi difference, which is expected
between the ground level of Ft.~Sumner, New Mexico and the air pressure at
altitude of 40 km, the additional offset was measured to be ($\Delta$az,
$\Delta$el) = (+12.1$'$,+13.9$'$), so that the total boresight offset of
\pe2\  relative to the star camera is (17.2$'$, 52.9$'$). The correction
for this boresight offset is applied in reconstructing X-ray sky images
from the \pe2\  observations below.


\begin{figure} 
\includegraphics*[width=0.49\textwidth]{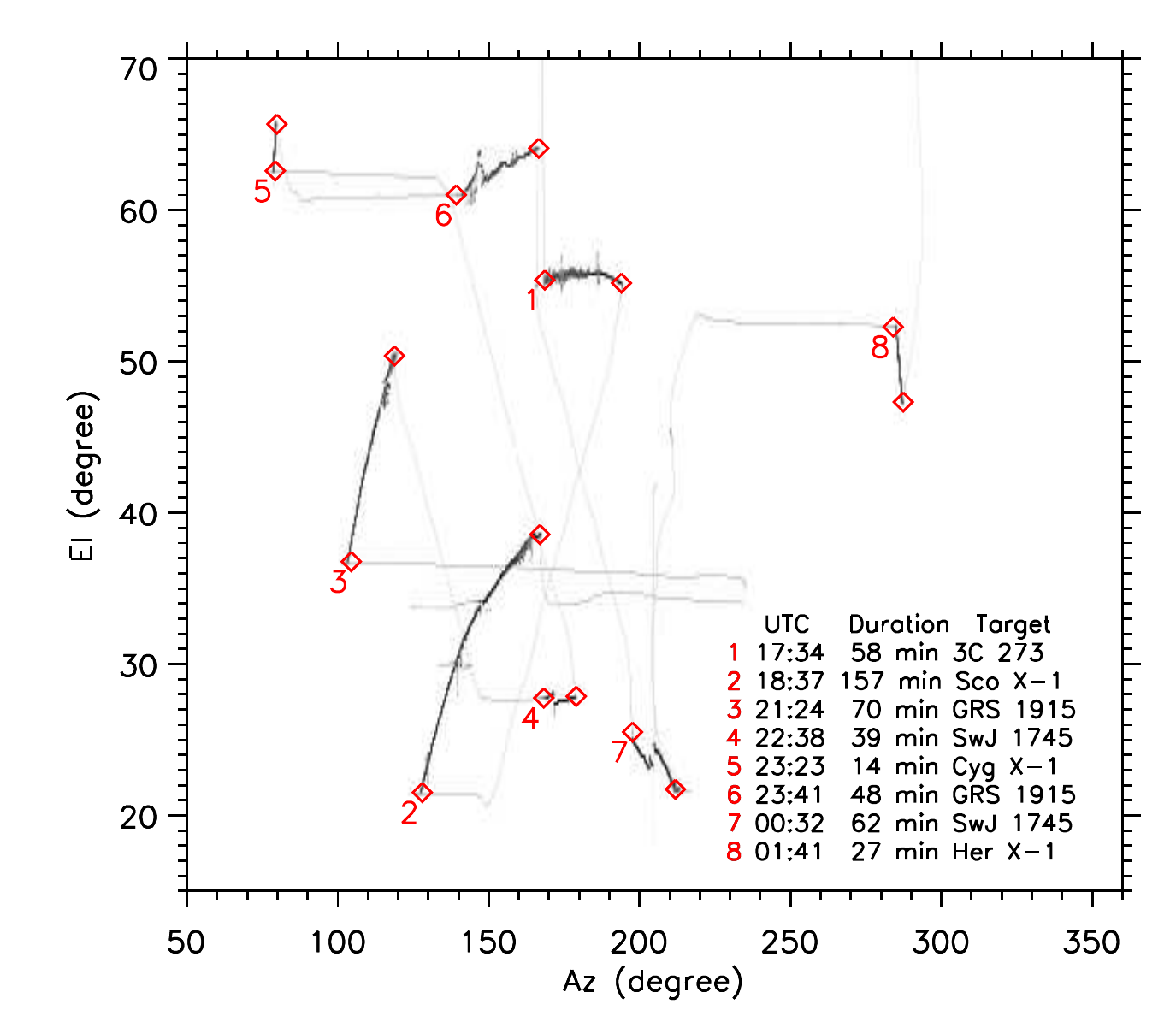}
\caption{Pointing history of the \pe\ telescopes during the 2012 flight in azimuth and elevation.
The number indicates the sequence of the observations. 
}
\label{f:sc}
\end{figure}

\begin{figure} 
\includegraphics*[width=0.45\textwidth]{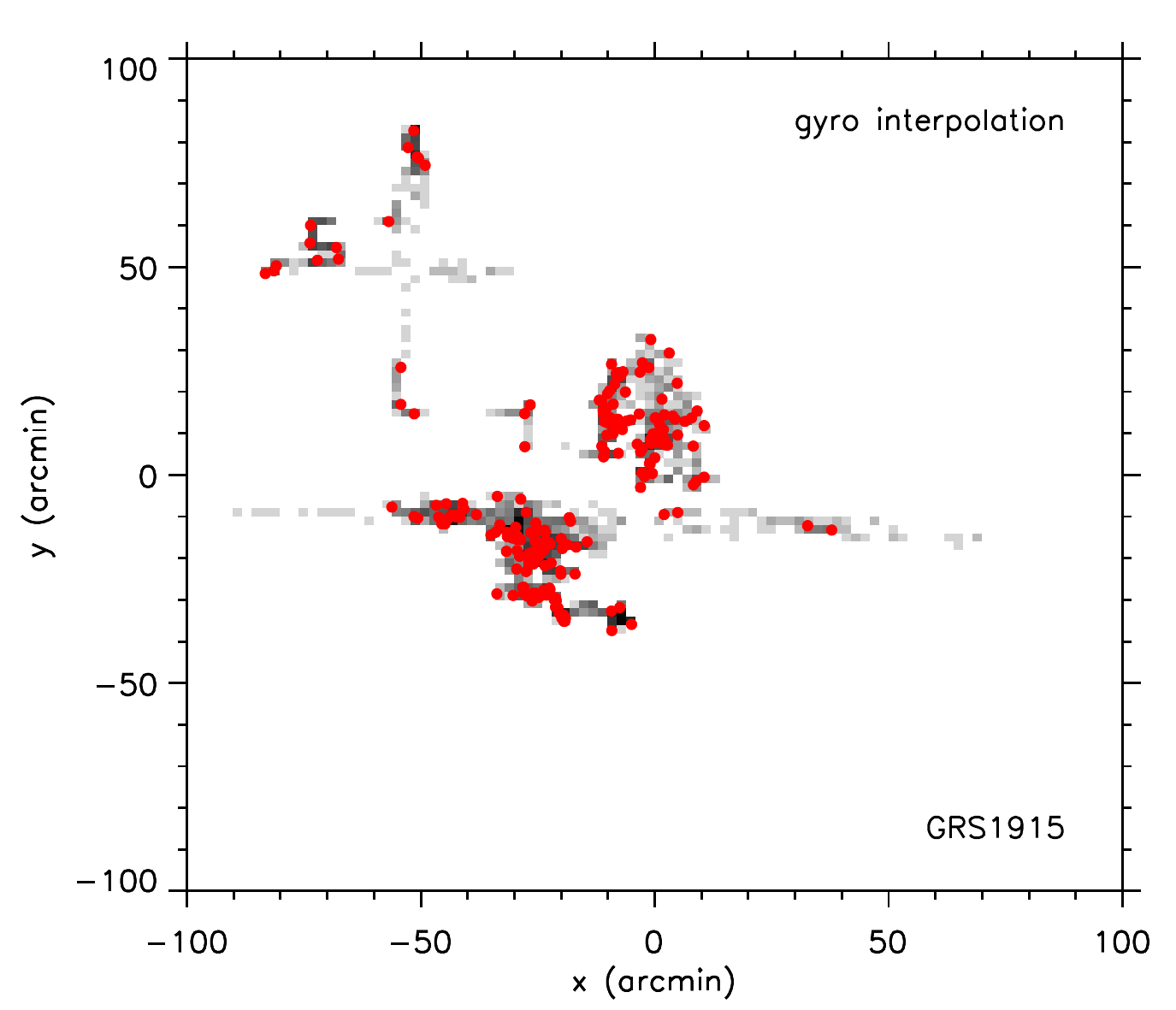}
\includegraphics*[width=0.45\textwidth]{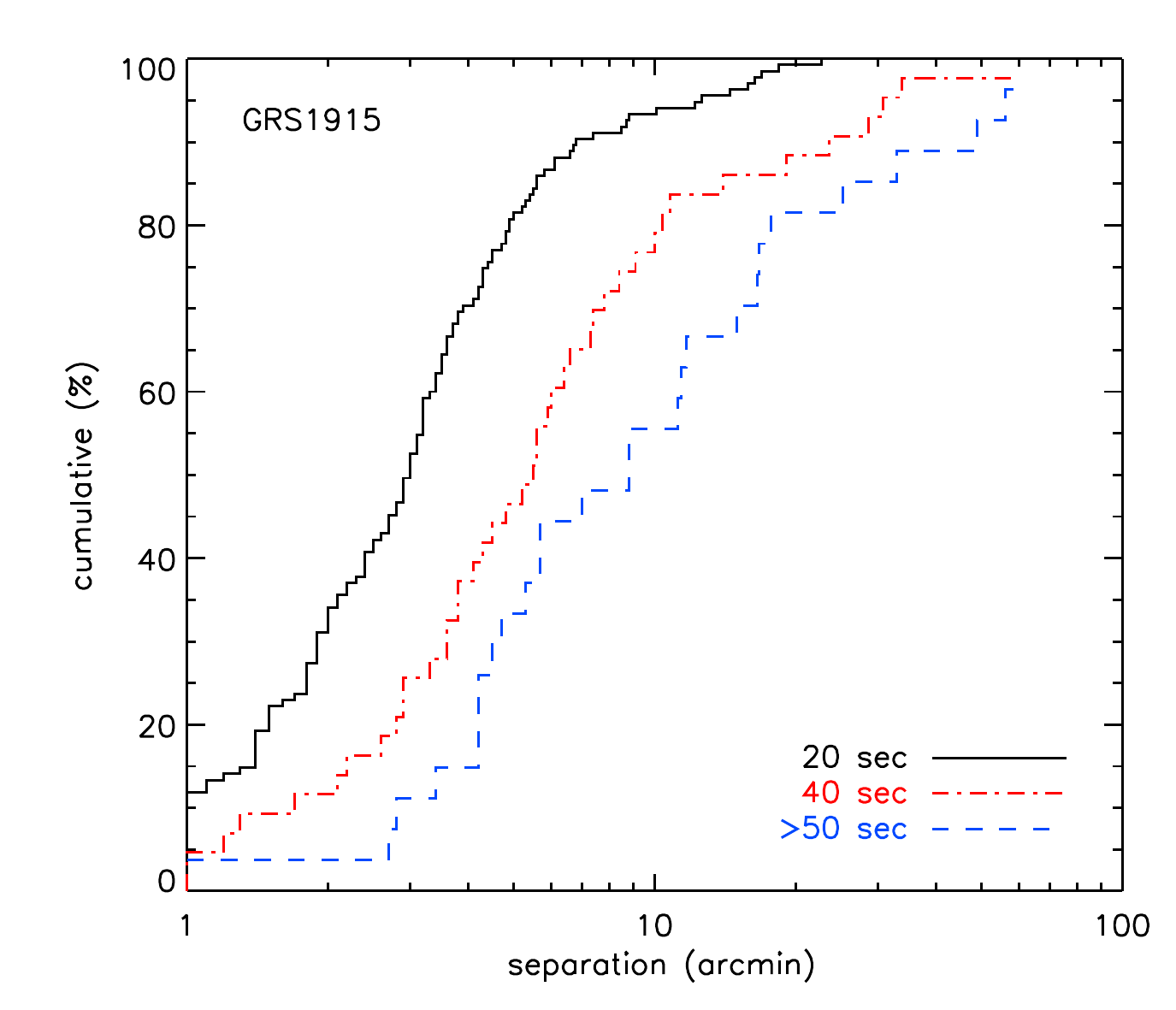}
\includegraphics*[width=0.45\textwidth]{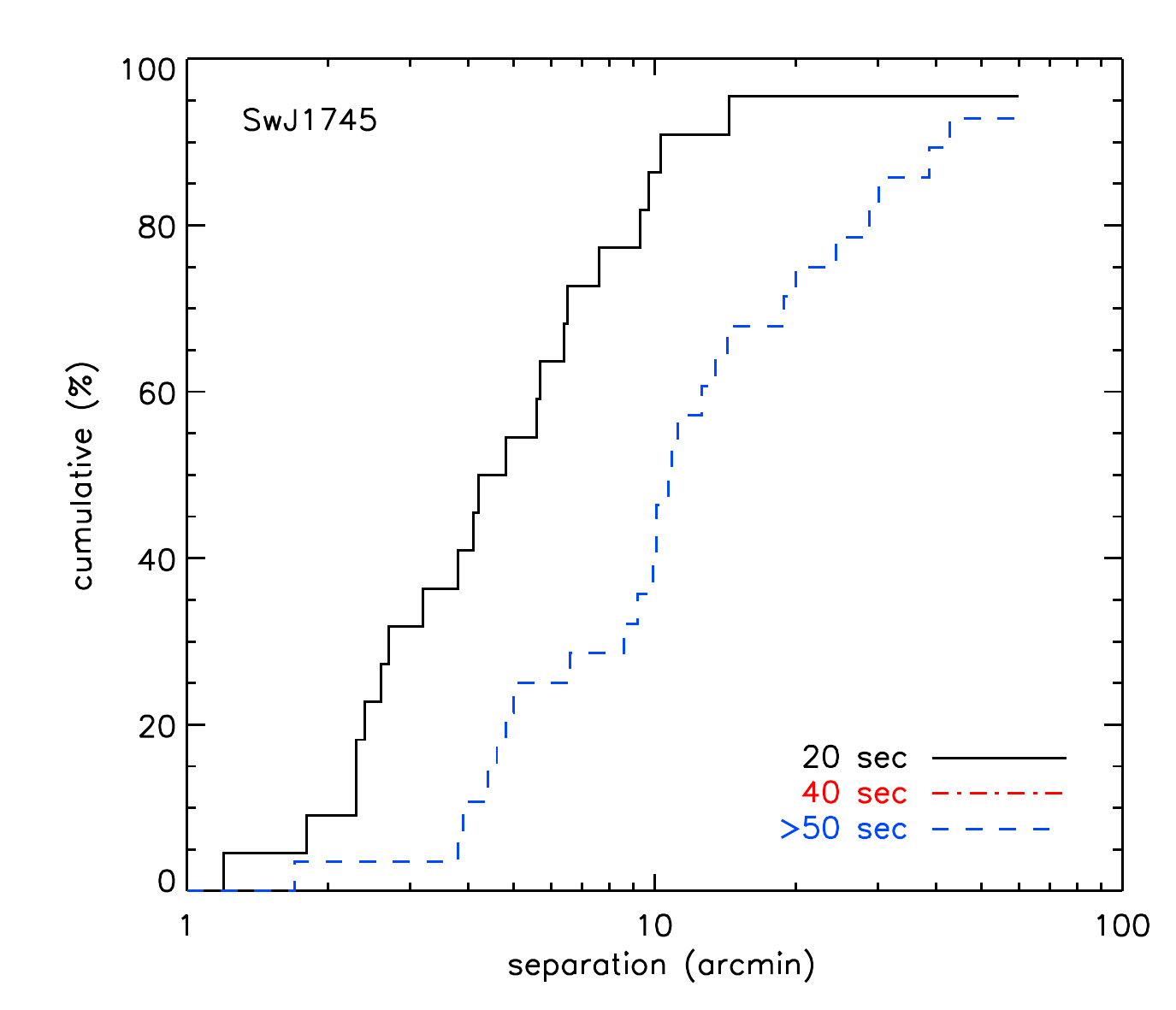}
\caption{
(Top) Positions of GRS~1915+105 relative to
the aimpoint based on the 206 star camera images (red dots) and 
Gyroscope interpreted estimates in between the star camera images (gray scale).
(Middle) Accumulative histograms of changes in pointing direction between
two successive star camera images during the observations of GRS1915+105
and (Bottom) Swift~J1745.1--2624.
}
\label{f:hoff}
\end{figure}

Figure~\ref{f:sc} shows the pointing history of the \pe\ telescopes during
the 2012 flight in azimuth and elevation.  We observed 6 X-ray sources
with 8 separate pointings that lasted 15 to 160 mins each.
While the \pe2 telescope performed well during the flight \citep{Hong13},
the pointing guidance system encountered a few issues that prevented
the telescopes from locking on each target. The top panel in
Figure~\ref{f:hoff} shows a scatter plot of the target positions
relative to the pointing direction retrieved from the 206
star camera images during the observation of GRS 1915+105.  The gray
tracks are interpolated points by the onboard Gyroscope in between two
successive star camera images.  While the targets were mostly in the
central
FoV during the observations as seen in Figure~\ref{f:hoff}, they were
constantly drifting beyond the angular resolution of the telescope 
at various speeds.  

We have to rely on the star camera images to recalculate
the precise pointing direction and apply the subsequent corrections.
Unfortunately, the pointing software was not designed to store all the
star camera images, and the number of the usable star camera images
saved during the flight were limited to 20 during the observation of Sco~X-1, 50 for 
Swift~J1745.1--2624
and 206 for GRS~1915+105.  
The bottom two panels in Figure~\ref{f:hoff} show accumulative histograms
of the changes in pointing direction between two successive star camera
images during the observations of GRS1915+105 (middle) and Swift~J1745.1--2624
(bottom).  The shortest interval between two star camera images was 20
sec and the next 40 sec apart. In 20 and 40 sec the pointing system
drifts more than 3\arcmin\ and 5\arcmin\ for a half of the time, respectively.
Although the pointing system became more stable by the second
observation of GRS~1915+105, the movement between two successive star camera
images is often larger than the angular resolution (4.8$'$).

This indicates that one cannot rely on the aspect information beyond $\pm$
10 sec within each star camera image. Unfortunately this severely limits
the data unusable for X-ray imaging. For instance, in the
case of GRS~1915+105, the relatively good time interval (GTI) is about 69 mins
out of the 118 min observation. For the rest of sources, the GTIs range
from about 7 to 14 mins total per source.  

\subsection{Sensitivity Estimate}\label{s:sens}

\begin{figure} \begin{center}
\includegraphics*[width=0.49\textwidth]{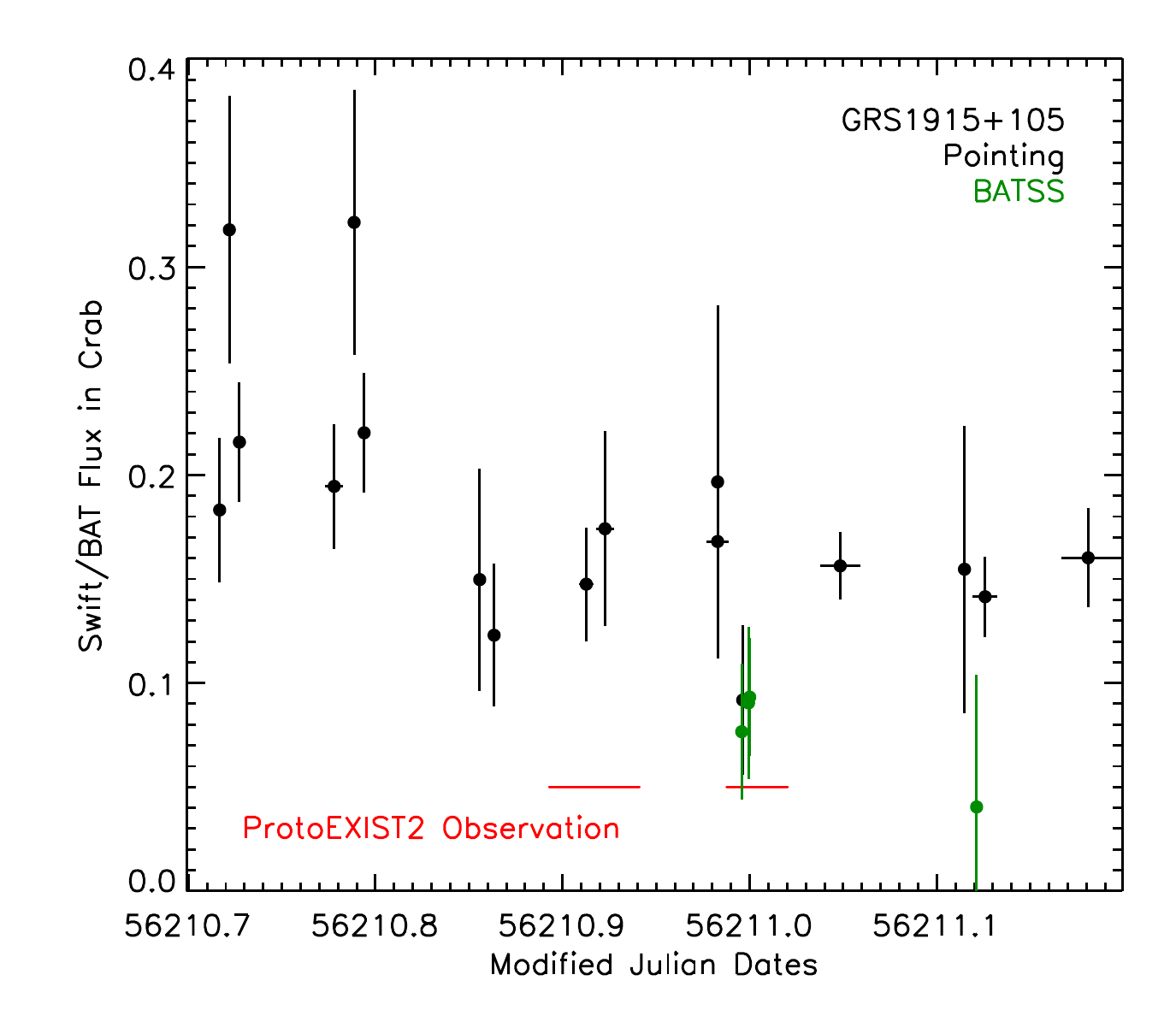}
\caption{X-ray fluxes of GRS1915+105 in 15--50 keV measured by the \swift/BAT pointing
and BAT slew survey (BATSS). The red lines indicate the
intervals of the \pe2 observations of the source. The 
detection limit of \pe2 is about 450 mCrab (5$\sigma$) for a 4000 s
(GTI) observation (see Section \ref{s:sens}).
}
\label{f:grs1915}
\end{center}
\end{figure}

We mainly focus on the observation of GRS 1915+105, which 
was conducted during a relatively stable pointing period and produced
the largest number of the star camera images.
Figure~\ref{f:grs1915} shows the lightcurve of GRS 1915+105 measured by
the \swift/BAT pointing and slew survey (BATSS) \citep{Copete12} around the \pe2\  observations
of the source.  A longer term lightcurve of the
source shows that the 15--50 keV flux of GRS~1915+105 was $\sim$ 400
mCrab weeks before our observation but fluctuated down to
$\sim$ 100--150 mCrab during our observation.

For an about 70 min observation of a 150 mCrab source, we expect 
1400 photons on the detector in 30--100 keV
after accounting for the atmospheric attenuation (by a factor of 3 on average).  We observed
about 1.1 cps per detector unit in 30--100 keV, so the total background
is about 270000 counts. For an ideal system with perfect pointings
and pure Poisson-driven background, we expect a signal-to-noise ratio (SNR)
of 2.7. The finite mask to detector pixel ratio (1.8) introduces
an imaging factor of 0.8 \citep{Skinner08}, causing an additional loss
in SNR by 20\%. Assuming a relatively moderate
aspect blurring of 1$'$ (likely an underestimate given the performance of
the pointing system), we expect another reduction in SNR by $\sim$ 20\%.
We also expect that the non-uniformity in the background
reduced the SNR by another $\sim$ 10\% when untreated (see Section \ref{s:sbc}).
Therefore a more realistic SNR expected from the
source is $\lesssim$ 1.7.   The sensitivity limit for a 5$\sigma$
detection in a 4000 sec observation is $\gtrsim$
450 mCrab.  
The limited GTIs due to relatively unstable pointings
and the unexpectedly low hard X-ray fluxes of the observed sources
ultimately hampered our ability to detect the sources.

According to \swift, the 15--50 keV X-ray fluxes of Swift J1745.1--2624 and Sco X-1
were about 500 mCrab and 1.5 Crab, respectively, during the time of the
\pe2 observations. However, for Swift J1745.1--2624, only 50 star camera images were
saved and the pointing system was more unstable during the observation
of Swift J1745.1--2624 by about a factor of 2 compared to the GRS 1915+105 as seen in
Figure~\ref{f:hoff}.  The low elevation of the source would have also
increased the atmospheric attenuation (Figure~\ref{f:sc}).
In the case of Sco X-1, only 50 star camera images are available
and its X-ray emission is dominantly soft, mostly below
30 keV, which is likely attenuated by the atmospheric absorption at
the altitude of our flight.

\begin{figure*} \begin{center}
(a) \hspace{-2ex}\raisebox{2.7ex-\height}{\includegraphics*[width=0.45\textwidth]{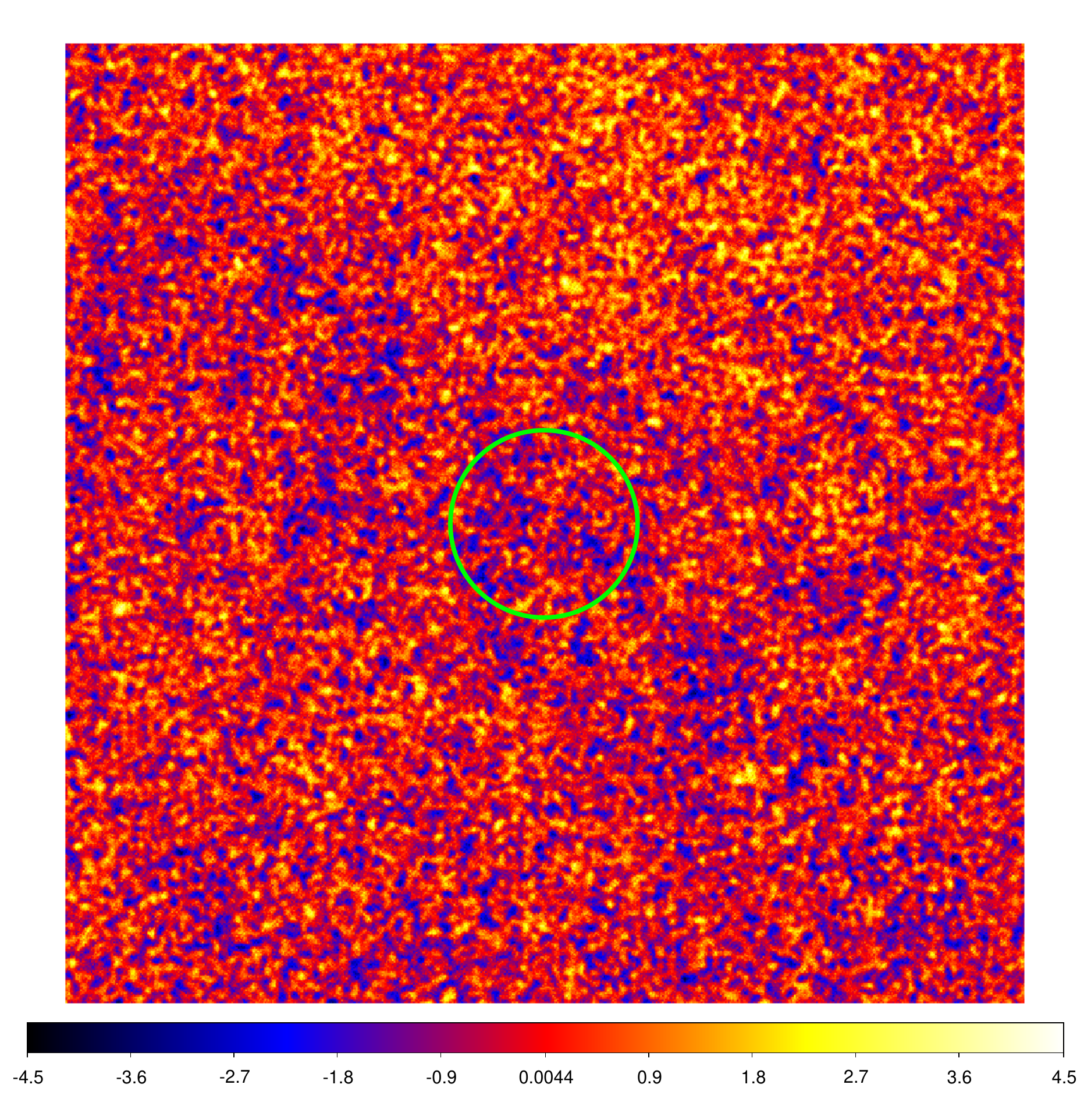}}
(b) \hspace{-2ex}\raisebox{2.7ex-\height}{\includegraphics*[width=0.45\textwidth]{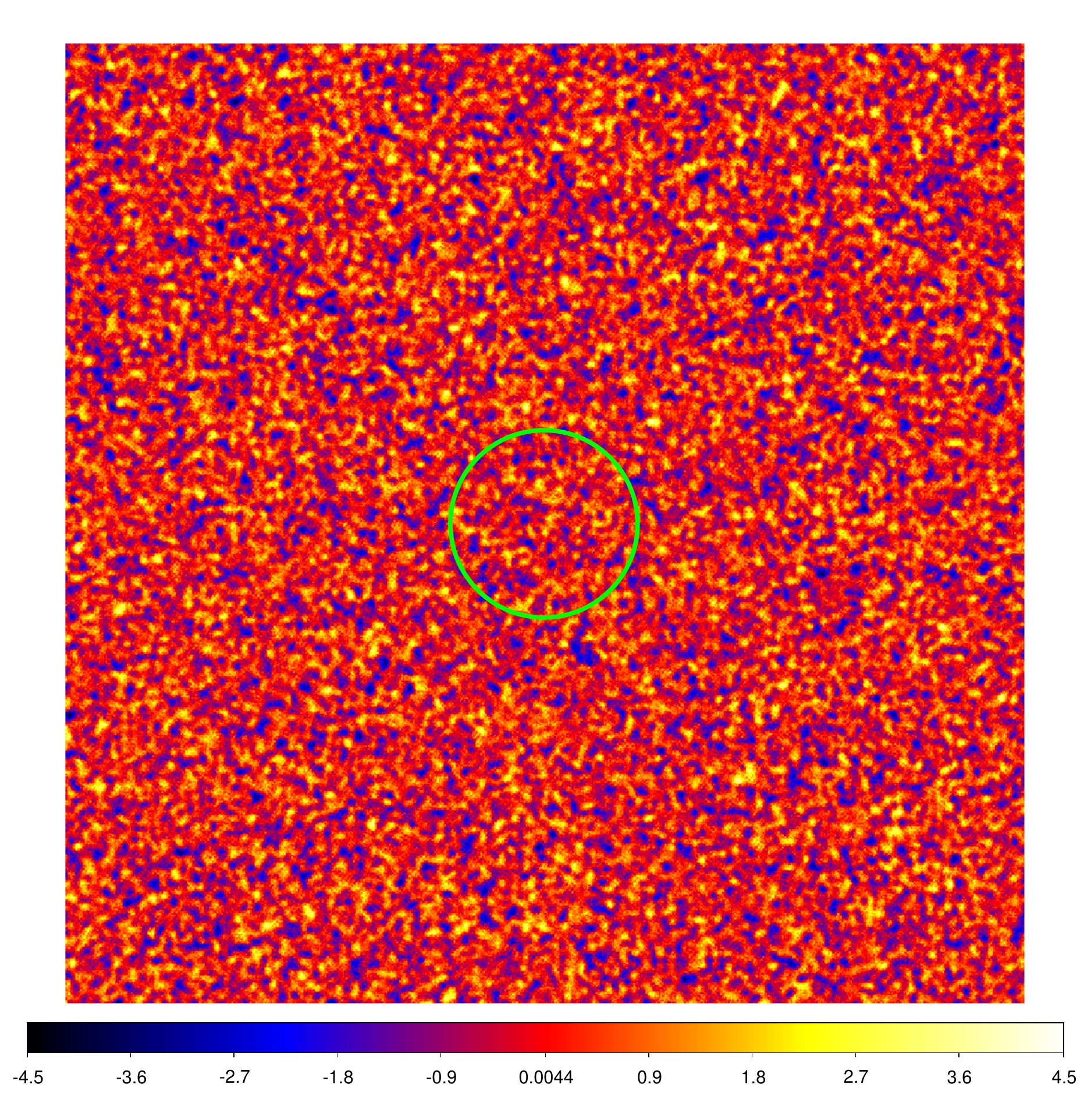}} \\
(c) \hspace{-2ex}\raisebox{2.7ex-\height}{\includegraphics*[width=0.45\textwidth]{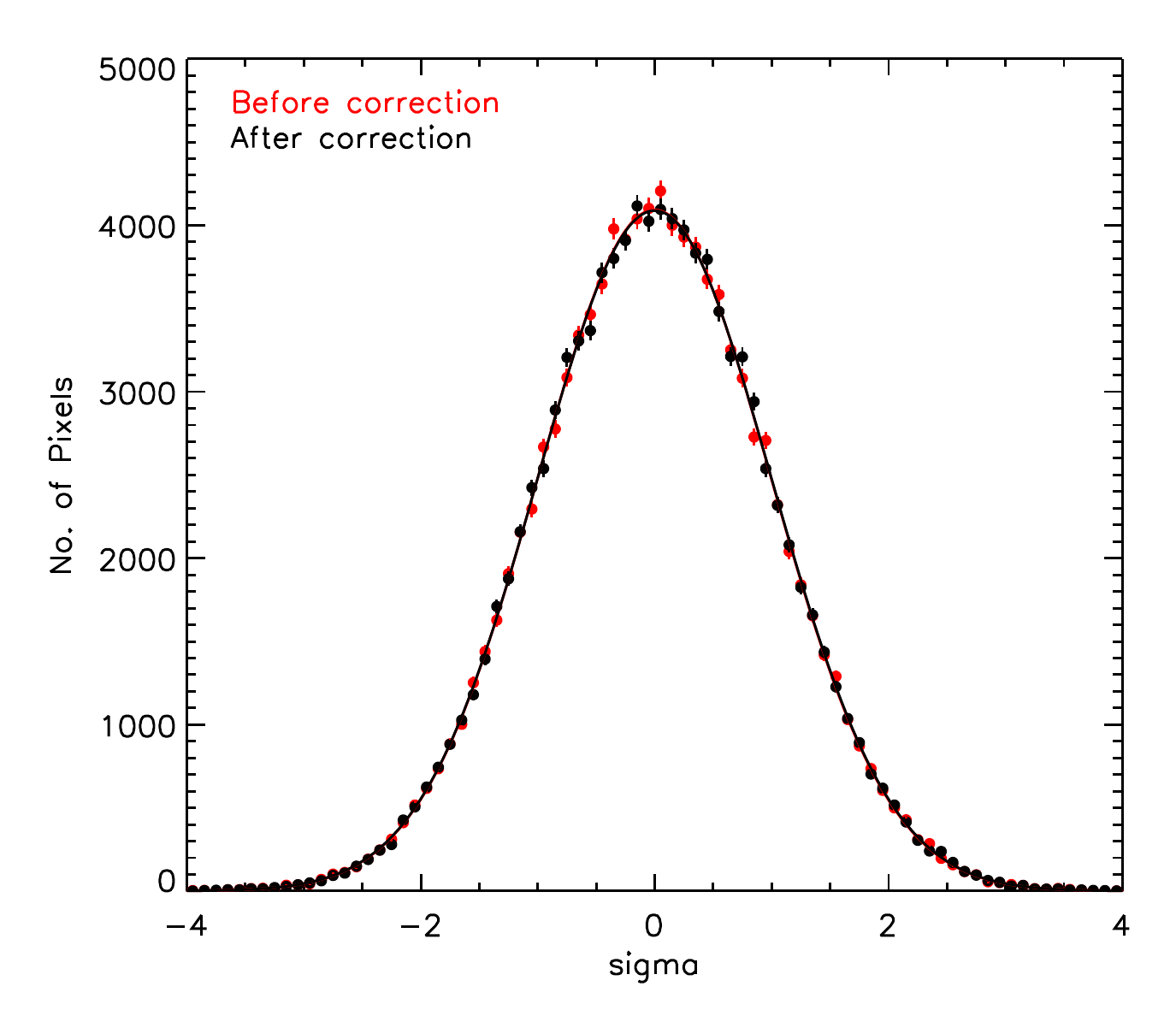}}
(d) \hspace{-2ex}\raisebox{2.7ex-\height}{\includegraphics*[width=0.45\textwidth]{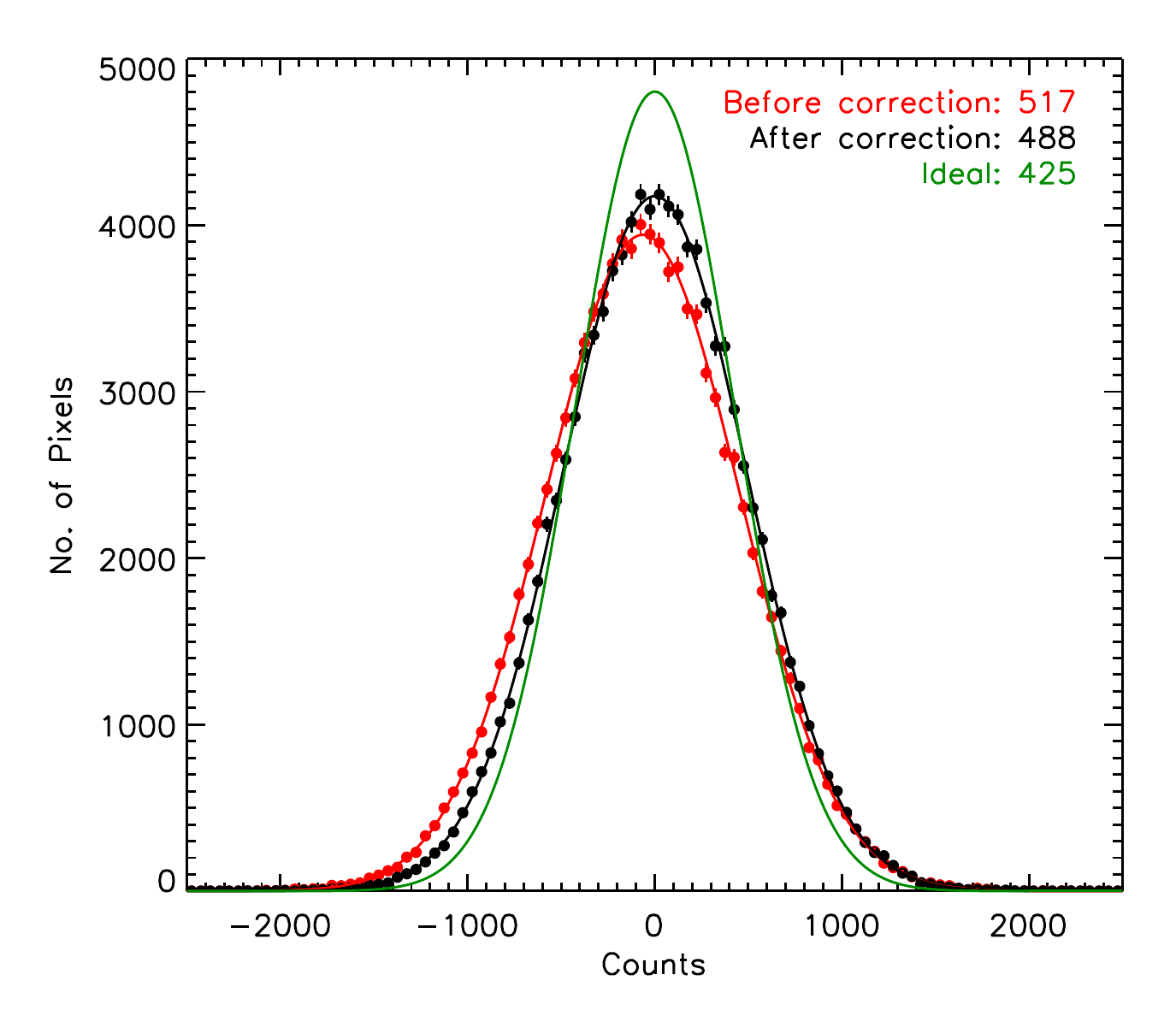}}
\caption{Reconstructed sky images of significance (30--100 keV) of an
about 10\Deg $\times$ 10\Deg region around GRS1915+105 reconstructed from
combining the \pe2 data of 20 sec intervals around each star camera
image (206 total): (a) without background subtraction, (b) with a
self-background correction scheme (Section \ref{s:sbc}). The circle
indicates a 1 deg radius of GRS1915+105. Pixel distributions of (c) the
significances and (d) raw counts with 2\arcmin\  sampling frequency (the rms
values are shown in the label).
The background correction scheme reduces the rms fluctuation of the
raw counts by about 5--7\%, which is $\sim$ 13\% larger than the ideal
case (green).
}
\label{f:xray}
\end{center}
\end{figure*}

Figure~\ref{f:xray} shows \pe2 X-ray images of a 10 deg $\times$ 10 deg
region around GRS 1915+105. We extracted the data in
a 20 sec interval around each of the 206 star camera images,
and combined them with proper boresight and aspect corrections using the
world coordinate system (WCS) of the star camera images.
The reconstructed sky image on the left panel is without any
background treatment and on the right with the background treatment
described in the next section.  There is no significant source in the
FoV as expected from the SNR estimates in the above.  Although the
noise distribution in the sky image
is well fitted by a Gaussian distribution, the sky image created without
any background treatment shows large scale structures
(e.g.~diagonal and circular ridges), indicating the
effects of the non-uniform background in the detector. 
In 20 sec, the average counts in each
pixel of the detector is about 0.02 counts, so the non-uniform
pattern of the background is not apparent in each individual detector
plane image.




\section{Self-Correcting Background Scheme}\label{s:sbc}

\begin{figure} \begin{center}
\includegraphics*[width=0.49\textwidth,trim=1cm 0cm 0cm 0cm]{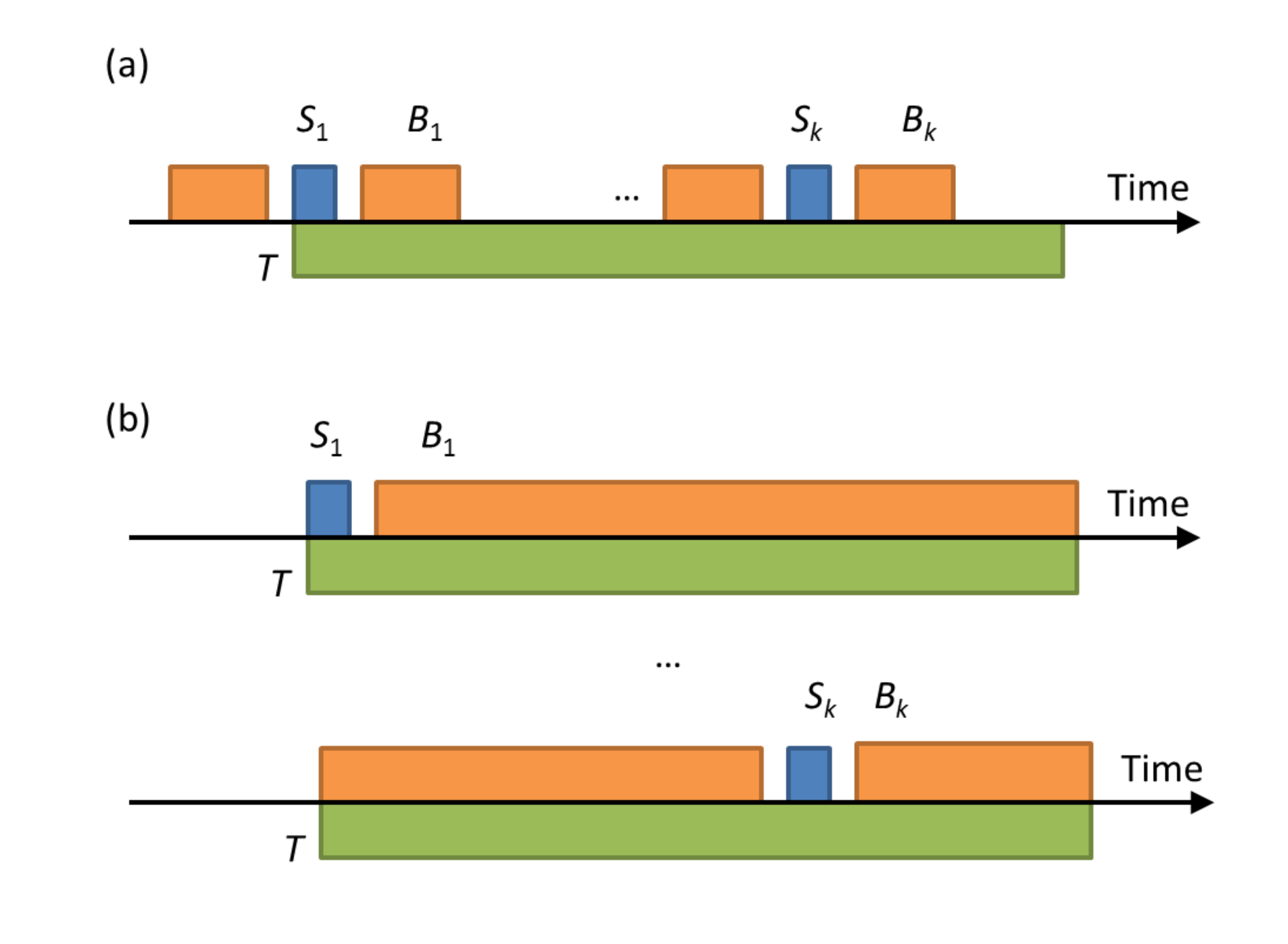}
\caption{Illustration of the data selection ($B_k$) for background modeling for each segment ($S_k$)
(a) sliding window selection (b) anti-segment selection.
}
\label{f:segsel}
\end{center}
\end{figure}



%

\begin{figure*} \begin{center}
\includegraphics*[width=0.45\textwidth]{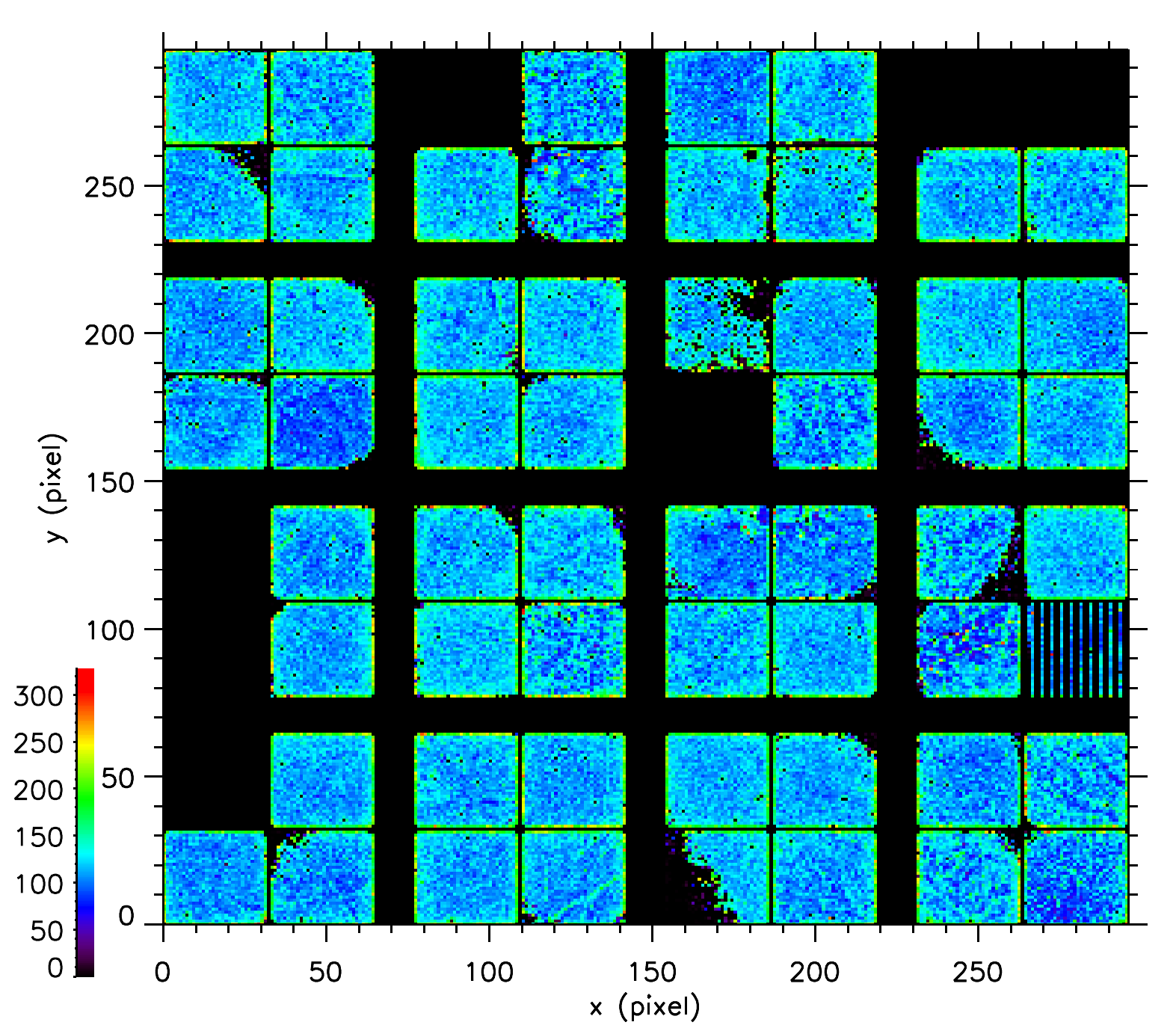}
\includegraphics*[width=0.45\textwidth]{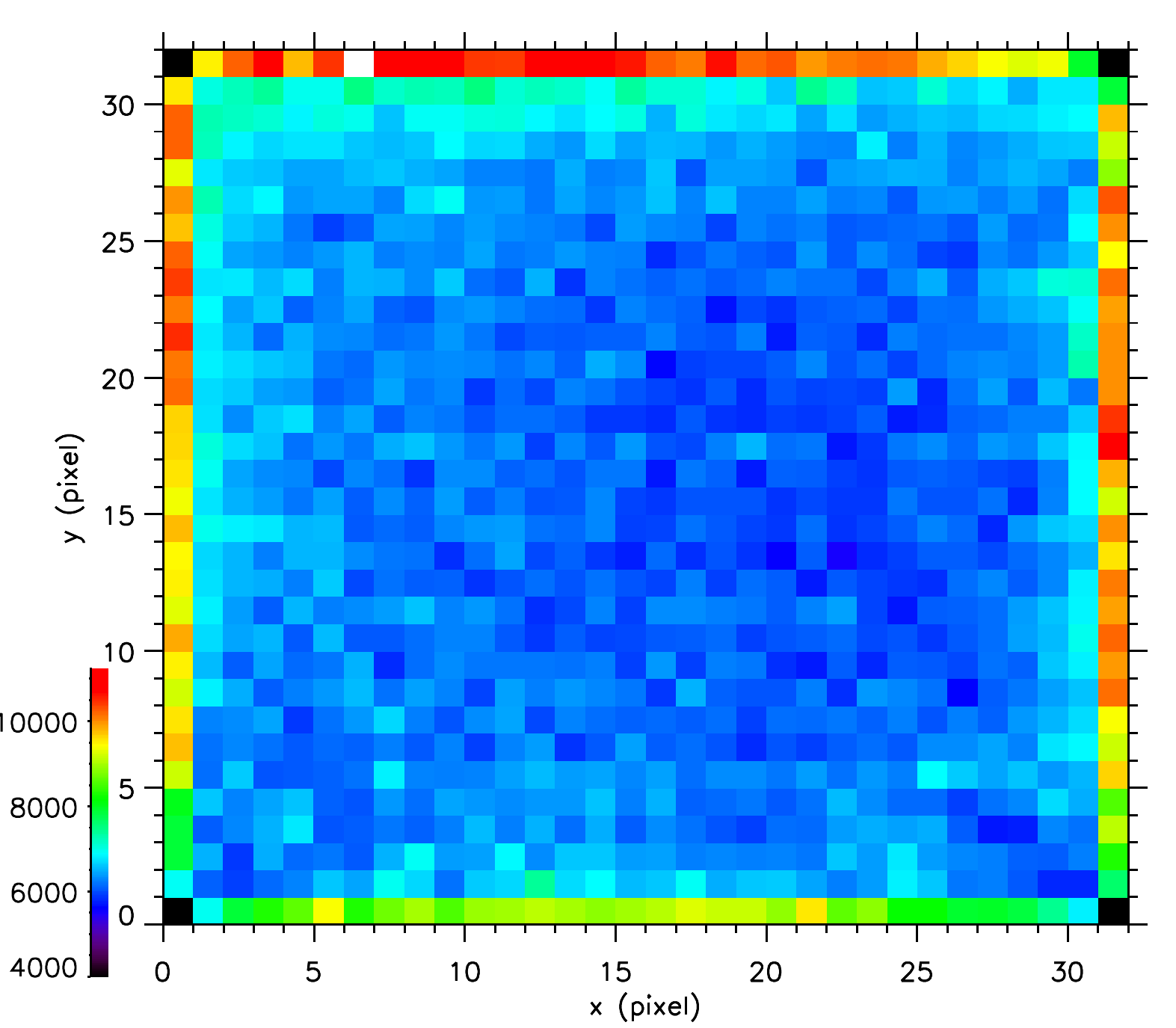}
\caption{Event distributions in the detector plane (left) and the same
but folded onto a single detector unit (right) \citep{Hong13}.
}
\label{f:detimg}
\end{center}
\end{figure*}

Figure~\ref{f:detimg} shows the distribution of X-ray events in the detector 
during the flight. The gaps between CZT crystals including
additional space the wirebonds used for control and readout of the
\nustar ASICs expose the side of each crystal to background radiation. As a result, 
edge pixels experience higher background counts than inner pixels.
The edge pixel background enhancement is
more clearly visible when the event distribution of each unit is stacked together
(the right panel in the figure).  
Since the size of the gaps varies
across the detector plane, each unit will have a different enhancement in
the edge pixels.  Besides the edge pixel background enhancement, there is also a more
gradual trend of count increase toward the edges.  The overall
view of the event distribution in the detector (the left panel) also
reveals other background patterns that are not originated from the
coding pattern of the mask. These non-uniform background patterns are
the source of large scale structures in reconstructed sky images (the
left panel in Figure~\ref{f:xray}).

A common way to handle non-uniform background in 
coded-aperture telescopes is first establish a model of the background
pattern from measurements and then subtract the properly scaled background model from
the data set.  For instance, in \swift/BAT,
the detector plane image is `cleaned' with a 14-element background
model \citep{Krimm13}.

During the 2012 flight, \pe2 recorded about 1 counts ks\sS{-1}
pix\sS{-1} in the 30--100 keV band after excluding X-ray events from
the onboard \sS{241}Am source. Over an hour observation, each pixel would
accumulate about 3 counts on average, which is 
too small to build  a reliable analytic model for the
background. Using a longer interval for background modeling 
would enable higher count statistics but it is not favorable
since the background pattern often varies with time.

Alternatively one could try smoothing the event distribution 
to better identify the background pattern under the
assumption that the background pattern does not have high spatial
frequencies. However, as seen in 
the edge pixel enhancement due to detector gaps
(Figure~\ref{f:detimg}), it is not unusual to have high frequency
noise fluctuations. Since bright sources will also generate high
frequency fluctuations in the detector plane, it often requires complex
multi-component background modeling with proper cleaning of bright
sources to properly handle the systematics in the detector plane
\citep[e.g.,][]{Krimm13}.


In order to combat non-uniformity in the detector response in general,
a continuous scan or slew motion instead of a fixed pointing has been proposed
\citep{Grindlay04, Copete12}.  During a scan,  the relative
position of each detector pixel with respect to a given sky pixel changes
continuously, which allows any detector-coordinate
dependent systematic errors to average out to some degree depending on
how thoroughly the scanning operation covers the given FoV.  In fact, scanning
operation can be viewed as an extreme version of dithering motions,
which are commonly used to reduce the systematics for both focusing and non-focusing telescopes.

In \swift/BAT, the roll angle is varied within $\pm$~1~deg from an
orbit to another to prevent the systematics errors from accumulating.
The BAT slew survey (BATSS) takes advantage of the self-correcting
feature in slew motions \citep{Copete12}.  In BATSS, given the high
speed slews performed by \swift, one sky image is generated every 0.2
sec in order to minimize blurring, which is later combined to a single
stacked image for source detection.  Counts in each pixel of each 0.2 s image are relatively small
and the pre-calculated background model may be not suitable for the rapid
change of the telescope orientation during the slew.  Nonetheless, the
images generated by BATSS with no background handling procedure show an
improvement relative to the images cleaned by a sophisticated background
model from pointed observations of the equivalent duration
\citep{Copete12}.  However,
similarly to  X-ray images generated from the \pe2 data, the sky images
from BATSS show large scale fluctuations due to
lack of any treatment on the non-uniform background.


Figure~\ref{f:segsel} illustrates our new approach to auto-correct
non-uniformity in the background under scanning or dithering motions
while (unknown) sources of interests are in the FoV.
The total duration of the observation is shown in the green interval
($T$). First we divide the data into multiple segments, where each
segment is well separated in terms of either pointing direction or
roll angles more than the angular resolution. The figure shows, for the $k$-th segment, the interval of
each image ($S_k$) in blue and the background data set ($B_k$) in
orange.  In the case of Figure~\ref{f:segsel}(a), each interval used for
the background is chosen to surround the segment of the
interest ($k$): ideally the interval ($B_k$) should be long enough to have
much higher photon counts than the same in $S_k$ (perhaps by a factor of 10 or more),
but not long enough to allow any significant changes in the background
pattern.  
If the background pattern has not changed significantly during the observation,
the entire observation excluding each segment and its immediate neighbors can be used for 
the background interval for the segment as shown in Figure~\ref{f:segsel}(b).

Assuming there is not enough photon statistics in any of $B_k$ to create
a reliable analytic model, for each segment ($S_k$) we simply subtract the scaled
(by $S_k$/$B_k$) raw counts in $B_k$ from $S_k$ (see Section \ref{s:tm} to
handle the background without subtraction), and generate a background
subtracted detector image. Then we combine the image of each segment with
a proper aspect correction to make a final stacked image for source detection.

\begin{figure} \begin{center}
\includegraphics*[width=0.49\textwidth]{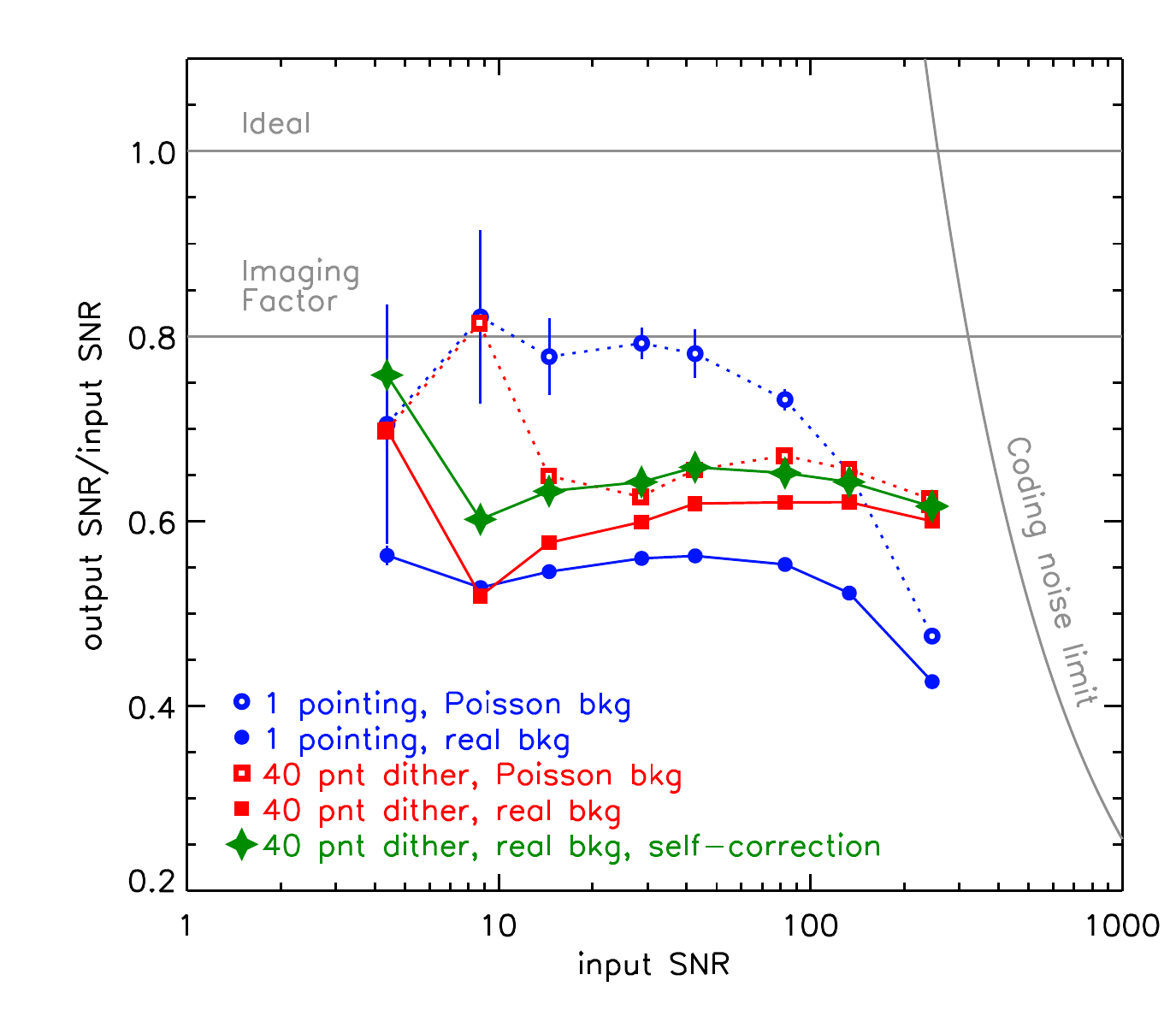}
\caption{Input and measured SNR ratio for various simulated sources
with purely Poisson backgrounds (dotted lines with open symbols) and
the real \pe2 background (solid lines with closed symbols) under a
single pointing observations (circles) and 40 point dithering operation
(squares and diamonds). The error bars are drawn from 10 cases of
single pointing observations with Poisson background for illustration.
}
\label{f:snr}
\end{center}
\end{figure}

If changes in the pointing or roll angles between the segments
are equivalent to or smaller than the angular resolution of the telescope,
this procedure will eliminate the source signal as well as the background
fluctuations since in essence it is a self-subtraction procedure.  However,
if each segment is different in pointing direction or roll angle larger
than the angular resolution, this procedure will eliminate the background
pattern tied to the detector coordinates within the limit of the given
photon statistics while allowing the source counts to accumulate at the
right sky position.

The right panel in Figure~\ref{f:xray} shows an X-sky image 
generated with the background subtraction following the recipe in
Figure~\ref{f:segsel} (b), which clearly shows a drastic reduction in the
large scale variations, compared to the case without any background
subtraction (a).  In addition, while the overall noise distributions
of the sky images in Figure~\ref{f:xray} (a) and (b) 
follow Gaussian distributions as seen in Figure~\ref{f:xray}
(c), the noise distributions of the sky images in the raw count unit
in Figure~\ref{f:xray} (d) show that their noise spread is in fact 
larger than the spread expected from the pure statistical fluctuation 
of the observed counts (green). The large scale variations in
Figure~\ref{f:xray} (a) are also the cause of the offset in the
noise distribution (red) in Figure~\ref{f:xray}(d).  
These indicate that there is indeed the systematic-driven
contribution in Figure~\ref{f:xray} (a).  
The self-correcting background scheme reduces the rms of the noise by 5--7\%
in addition to largely removing the large scale variations.
Since our data did not have any detectable sources, in order to make
sure that the source signal under the self-correction scheme does not
get canceled out, we performed the same procedure with simulated sources.

Figure~\ref{f:snr} shows the ratio of the input and measured output SNR from
simulated sources with a wide range of input signals. For background,
we tested both pure Poisson statistics based backgrounds (dotted
lines with open symbols) and the \pe2 data (solid lines with solid
symbols).  Observations with fixed pointings (circles) are compared with
the 40 point dithering motions (squares and diamonds).  The results of
a fixed pointing with pure Poisson noise set an ideal limit
of the system, which is bounded by the imaging factor at low SNRs (80\%) \citep{Skinner08}
and by the coding noise limit at high SNRs.  
The coding noise is the mask pattern-induced noise \citep{Skinner08}.
The coding noise limit of
the output SNR is about 296 for \pe2.  In observations with fixed pointings,
the real background reduces the SNR down to 40-50\% (closed circles)
of the input SNR.

Dithering motions (red) alleviate the coding noise limit by mixing
different parts of the mask pattern in the detector image, and thus
outperforms pointed observations at high SNRs.  To simulate realistic
aspect corrections, we allowed about 1$'$ errors in simulated source
positions, so even with pure Poisson noise backgrounds (open squares)
the measured SNRs in the reconstructed images are lower than the input
SNRs by about 60--70\%.  With the real background (closed squares), the
output SNRs drop even more but not as much as the pointed observations,
which illustrates the benefits of dithering or scanning motions.
With the proposed treatment on the non-uniform background (green
diamonds), the output SNRs recover almost up to the pure Poisson noise
cases, which means about 3-10\% improvements relative to the cases of no
background handling. The improvements at low input SNRs are relatively
larger: e.g.~the proposed method improves the output SNR from 8.4 to 9.2.
The improvement varies with the scale of dithering or scanning motions and
the number of segments. If the number of segments is too small
(or $B_k$/$S_k$$\ll$10), the treatment will increase the rms of the sky
background although it may still reduce the large scale structures.

\section{Poisson Statistics based Detection Significance Map}\label{s:tm}


A common method to reconstruct sky images ($s_i$; $i$ represents the sky pixel index) for coded-aperture telescopes
is to cross-correlate detector plane images ($d_j$; $j$ represents the detector pixel index) with the mask
pattern ($M_{ij}$=1 for open and 0 for closed pixels). The cross correlation is often
performed through fast Fourier transformations (FFTs). In order to set the
baseline value of the sky image at zero and the source counts at the
right signal, the mask pattern is balanced by open fraction ($\rho$), where
the balanced pattern $N_{ij}$ is 1 for open pixels, and $\rho/(\rho-1)$
for closed pixels.  The detector plane image ($d_j$) is background ($b_j$)
subtracted  and masked out for dead pixels or inactive zones, if any,
before cross-correlation. For a random mask pattern, detector plane
images are `rebalanced' to have the total counts of zero using a constant ($c$), which suppresses
large scale structures bigger than the angular resolution. 
The `rebalancing' procedure
is often essential for point source detection, but it suppresses a large scale
variation even if its origin is celestial.
In summary, the reconstructed sky image ($s_i$)
is given by \begin{eqnarray}
s_i = N_{ij} \cdot (d_j - b_j - c), \label{e:con}
\end{eqnarray}
where $\Sigma (d_j - b_j - c)$ = 0.

The background subtraction, although it efficiently addresses
non-uniformity of the detector, may not be optimal in
handling non-negative
counts of Poisson statistics especially for characterizing the detection
significance of faint sources near the
detection threshold.  An alternative approach is to calculate the
probability of having more than the observed counts from a random
fluctuation of the background counts based on the Poisson statistics. 
For every sky position in the FoV, one can estimate the total observed
counts ($p_i$) by cross-correlating the raw mask
pattern and the raw detector plane image:  $p_i = M_{ij} \cdot d_j$.

\begin{figure*} \begin{center}
\includegraphics*[width=0.47\textwidth]{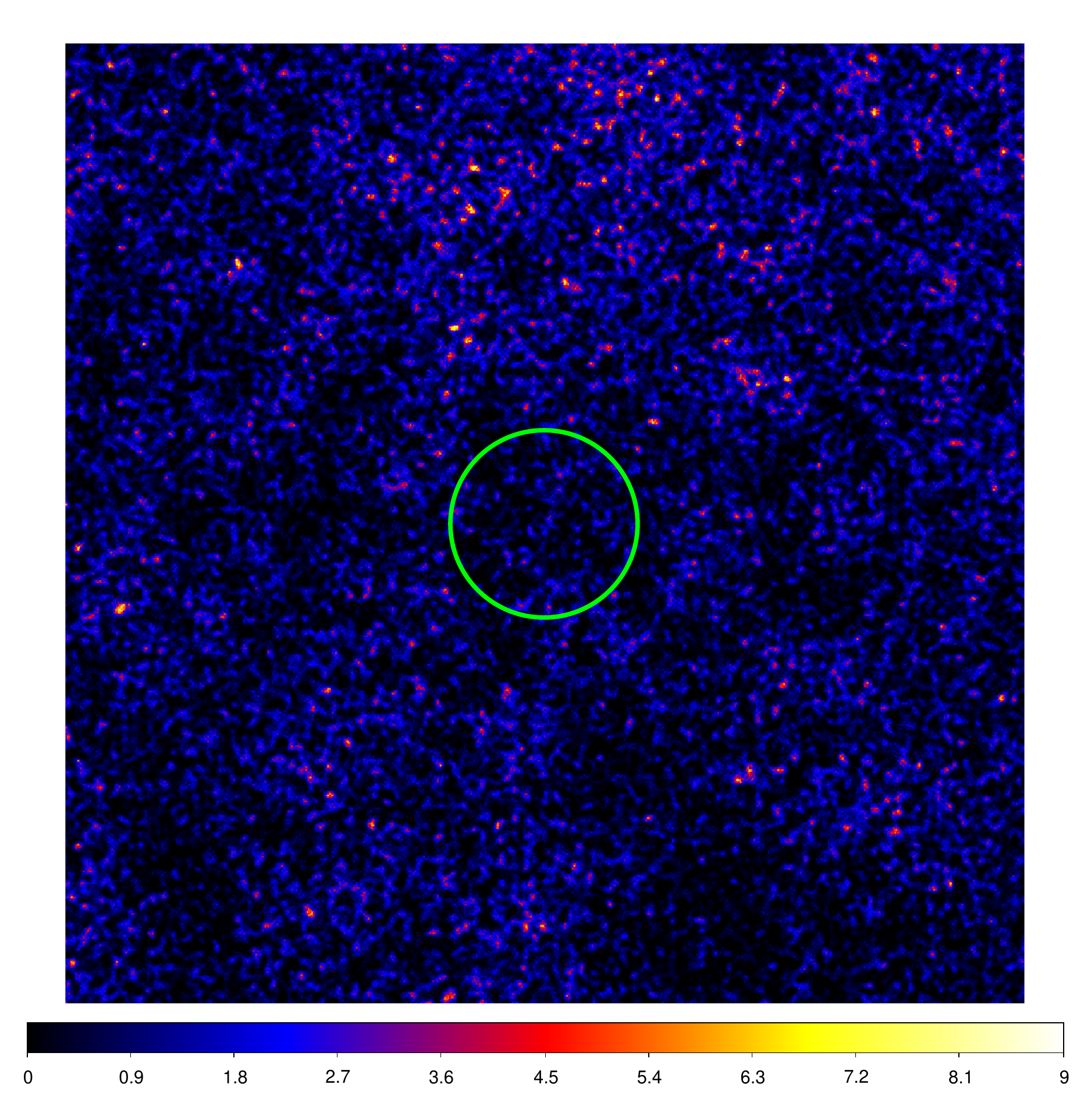}
\includegraphics*[width=0.47\textwidth]{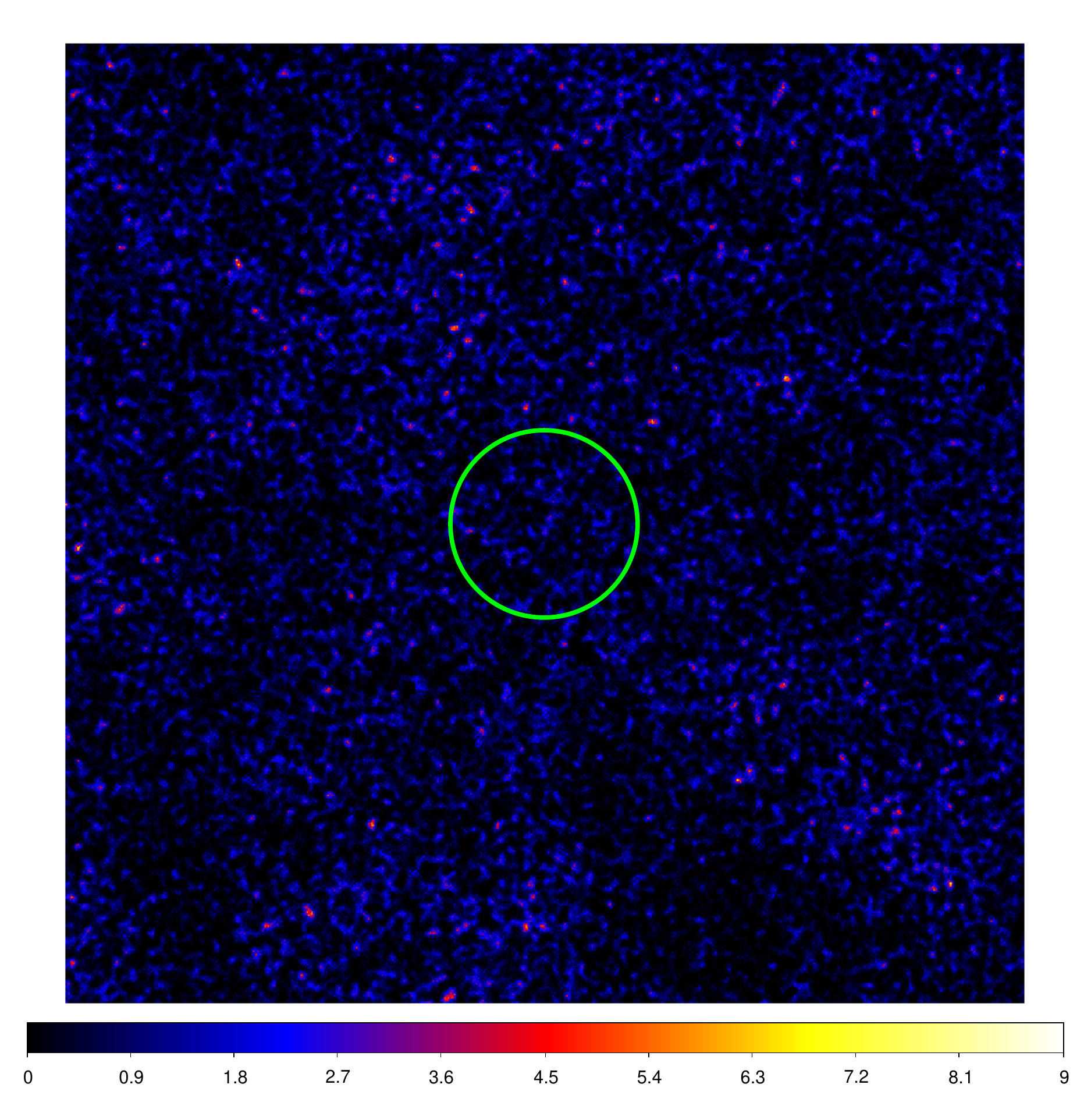}
\caption{Trial map of about 10 deg $\times$ 10 deg around GRS1915+105
from the \pe2 observation of the source using
20 sec intervals around each available star camera image (206 total):
(left) without any background treatment, (right) with a self-background
correction scheme (Section \ref{s:sbc}). The circle indicates a 1 deg radius
of GRS1915+105. The trial values on the right are about 10--100$\times$ lower.
The color scale is coded logarithmically with the trial numbers ($10^X$).
}
\label{f:tm}
\end{center}
\end{figure*}

The mask pattern ($M_{ij}$) consists of non-negative numbers representing 
open fraction ranging from zero for totally opaque elements to one for fully
open elements. The detector plane image also consists of non-negative raw counts.
For the estimation of the background counts,
we repeat the calculation using the background model.
\begin{eqnarray*}
	{q_i} = M_{ij} \cdot b_j, 
\end{eqnarray*}
Then for a given pixel ($i$) in the sky, the expected background ($r_i$) is
given  as 
\begin{eqnarray*}
	r_i 	& = &  \bar{p_i} q_i / \bar{q_i},  \mbox{where} \\
	\bar{p_i} & =&  \bar{M_{ij}} \cdot d_j,  \\
	\bar{q_i} & =&  \bar{M_{ij}} \cdot b_j.
\end{eqnarray*}
Note $\bar{M_{ij}}$ is an inverted mask pattern, 
which ranges zero for fully open elements to one for totally
opaque elements, but the non-imaging opaque elements remain zero.
Then, $\bar{p_j}$ and $\bar{q_j}$ represent the portion of the
counts that cannot come from the sky pixel ${j}$ among the coded detector
area for the sky position.

For the total observed counts ($p_i$) and the background estimate ($r_i$) at a given sky position ($i$),
the probability that the observed counts are purely due to a random fluctuation
of the background counts is given by a normalized incomplete gamma function \citep{Weisskopf07,Kashyap10}
\begin{eqnarray}
	P(>p_i) &=& \gamma(p_i +1 , r_i) \\
		& = & \frac{1}{\Gamma(p_i+1)} \int^{r_i}_0 e^{-t} t^{p_i} dt  \label{e:tm}
\end{eqnarray}
One can repeat the calculation for every sky position ($i$) to generate
the probability map.

In order to make a sky image in a conventional way where bright sources
have larger values, one can simply use an inverse of this probability map,
which is the number of trials required to generate the observed counts
from a random fluctuation. We call this map of required random trial numbers as a
`trial' map and it represents a significance map of source detection.
\citet{Hong16} first applied this trial map-based approach to point
source detection in X-ray images taken from \nustar, and in this paper
we extend the approach to coded-aperture imaging.
Since each pixel in a sky image of a coded-aperture
telescope represents one trial of source search, if the value in a
trial map greatly exceeds the total number of the pixels in the image,
it indicates a high chance of a real source (or a systematic artifact).
Detection threshold setting, therefore, is more straightforward in
trial maps generated by coded-aperture imaging than those by focusing telescopes,
where the threshold depends on the (potentially varying) size of the
point-spread function relative to the image.

A source in the partially coded FoV in a sky image generated by
Eq.~\ref{e:con} have a less signal than a source of the same strength in
the fully coded FoV, thus in order to reflect the source signal properly,
the image has to be renormalized by the partial coding fraction.  On the
other hand, such a normalization enhances the noise fluctuation in the
partial coded FoV.   When stacking multiple sky images generated by
Eq.~\ref{e:con}, they have to be weighted by the variance in order to
account for the proper coding fraction, where the error due to improper
handling of Poisson statistics can propagate and amplify.  In the case of
trial maps, they describe a chance of having a real source in any part of
the FoV regardless of the coding fraction. For stacking multiple images,
one simply accumulates the sky images of $p_i$ and $r_i$, then re-apply
Eq.~\ref{e:tm} to get the stacked new trial map.\footnote{In order to
`clean' bright sources (i.e., reduce the coding noise
associated with the bright sources), the sky images of all four quantities 
$p_j$, $q_j$, $\bar{p_j}$ and $\bar{q_j}$ have to be tracked. 
The cleaning procedure simply recalculates $r_j$ by treating the contribution
of the bright sources as a part of background, and thus it does not involve
any subtraction from $p_j$.}

Figure~\ref{f:tm} shows the trial maps of the same images in
Figure~\ref{f:xray}. The images are color coded logarithmically with
the trial numbers (10\sS{X}). Each image has about 60k independent pixels,
and the threshold for 0.1\% false detection is about $X$~$\sim$~7.8.
The background treatment described in Section \ref{s:sbc} reduces the large
scale structures and lowered the noise fluctuation by about 100 in
trial numbers, but the trial map on the right panel indicates there is
some residual non-uniformity in the image, which is not apparent
in Figure~\ref{f:xray}.

In terms of computational requirements, the total observed ($p_i$) and
background counts ($q_i$) can be estimated through FFTs. The latter can
be pre-calculated if the background model is known ahead and remains
unchanged. 

In coded-aperture imaging,  source detection often relies on a large
number of counts, but for high resolution imager aimed to detect
fleeting signals from transient sources, the total number of counts can
be relatively small, so that the suggested rigorous probabilistic
approach of source search is more appropriate.

\section{Conclusion and Future Work}\label{s:conclusion}

In future,  direct localization of transient sources like gamma-ray
bursts without assistance of secondary instruments will enable a wide
range of the time domain astrophysics.  To achieve this,  next generation
wide-field hard X-ray telescopes should be capable of sub 2\arcmin\ angular
resolution.  In high resolution coded-aperture telescopes, new challenges
arise in handling non-uniformity in the detector system due to low count
statistics per pixel. During a high altitude balloon flight in 2012, the
\pe2 telescope of 4.8\arcmin\ resolution collected about 3 counts per hr in
each detector pixel on average, which illustrates this new challenge.
Dithering or continuous scan as shown in BATSS alleviates the
effects of the systematics by automatically averaging out the
non-uniformity even without special treatment but large
scale variations still remain in their sky images, which can limit the 
detection and localization sensitivity.

We presented a method to improve the sensitivity of high
resolution coded-aperture systems by self-correcting the non-uniform
background of low statistics efficiently.  Combining simulated sources
with the real balloon flight data of the \pe2 telescope, which exhibits
pixel-dependent background variations, we demonstrated that the proposed
techniques can reduce the large scale variation dramatically and improve
the SNR by a few to 10\% depending on the input SNR.  We also proposed
a new method to estimate detection significance 
using a Poisson statistics based probabilistic approach
without relying on subtraction in background handling.
We plan to apply these techniques to the \swift/BAT data to evaluate the
improvements in a wide range of operating environments for further
optimization.

\section{Acknowledgment} 
This work was supported by NASA/APRA grant NNX14AD59G.

\bibliographystyle{elsart-num}

\begin{thebibliography}{10}


\bibitem[Gehrels et al.(2004)]{Gehrels04} 
	Gehrels, N., Chincarini, G., Giommi, P., et al.\ 2004, \apj, 611, 1005 

\bibitem[Winkler et al.(2003)]{Winkler03} 
	Winkler, C., Courvoisier, T.~J.-L., Di Cocco, G., et al.\ 2003, \aap, 411, L1 

\bibitem[Hong et al.(2013)]{Hong13} 
	Hong, J., Allen, B., Grindlay, J., et al.\ 2013, IEEE Transactions on Nuclear Science, 60, 4610 

\bibitem[Harrison et al.(2013)]{Harrison13} 
	Harrison, F.~A., Craig, W.~W., Christensen, F.~E., et al.\ 2013, \apj, 770, 103 

\bibitem[Skinner(2008)]{Skinner08} 
	Skinner, G.~K.\ 2008, \ao, 47, 2739 

\bibitem[Krimm et al.(2013)]{Krimm13} 
	Krimm, H.~A., Holland, S.~T., Corbet, R.~H.~D., et al.\ 2013, \apjs, 209, 14 

\bibitem[Grindlay \& Hong(2004)]{Grindlay04} 
	Grindlay, J.~E., \& Hong, J.\ 2004, \procspie, 5168, 402 

\bibitem[Hong et al.(2016)]{Hong16} 
	Hong, J., Mori, K., Hailey, C.~J., et al.\ 2016, \apj, 825, 132

\bibitem[Copete(2012)]{Copete12} 
	Copete, A.~J.\ 2012, Ph.D.~Thesis,  
	Harvard Univ., 316C

\bibitem[Weisskopf et al.(2007)]{Weisskopf07} 
	Weisskopf, M.~C., Wu, K., Trimble, V., et al.\ 2007, \apj, 657, 1026 

\bibitem[Kashyap et al.(2010)]{Kashyap10} 
	Kashyap, V.~L., van Dyk, D.~A., Connors, A., et al.\ 2010, \apj, 719, 900 

\end{thebibliography}

\end{document}